\documentclass[aps,preprint]{revtex4}

\usepackage[dvipdfm]{graphicx}
\usepackage{amsmath,amssymb}

\def\gtsim{\mathrel{\hbox{\raise0.2ex
  \hbox{$>$}\kern-0.75em\raise-0.9ex\hbox{$\sim$}}}}
\def\ltsim{\mathrel{\hbox{\raise0.2ex
  \hbox{$<$}\kern-0.75em\raise-0.9ex\hbox{$\sim$}}}}

\newcommand{\rs}[1]{_{\rm #1}} 
\newcommand{\rsu}[1]{^{\rm #1}} 
\newcommand{\bm}[1]{\mbox{\boldmath $#1$}} 
\newcommand{\dm}[1]{${\displaystyle #1 }$} 

\newcommand{\noscmax}{N_{\rm osc}^{\rm max}} 
\newcommand{\efree}[1]{\epsilon^{0}_{#1}} 
\newcommand{\cutoff}{f_{\rm c}} 

\newcommand{\spe}{\epsilon}  
\newcommand{\ecut}{\epsilon_{\rm cut}}
\newcommand{\rmax}{R_{\rm max}}
\newcommand{\ekmax}{\epsilon_{\rm kin}^{\rm max}}
\newcommand{\gtf}{g_{\!\!\phantom{.}_{\mbox{\tiny TF}}}}
\newcommand{\gobtf}{g_{\!\!\phantom{.}_{\mbox{\tiny OB}}}}
\newcommand{\ftf}{f_{\!\!\phantom{.}_{\mbox{\tiny TF}}}}
\newcommand{\fobtf}{f_{\!\!\phantom{.}_{\mbox{\tiny OB}}}}
\newcommand{\mean}[1]{\langle #1 \rangle}

\newcommand{\cJ}{{\cal J}}

\begin{document}

\title{ Method to circumvent the neutron gas problem\\
in the BCS treatment for nuclei far from stability }

\author{Toshiya Ono and Yoshifumi R. Shimizu}
\affiliation{Department of Physics, Graduate School of Science, Kyushu University, Fukuoka 812-8581, Japan}
\author{Naoki Tajima}
\affiliation{Department of Applied Physics, University of Fukui, 3-9-1 Bunkyo, Fukui 910-8507, Japan}
\author{Satoshi Takahara}
\affiliation{Kyorin University, School of Medicine, Mitaka, Tokyo 181-8611, Japan}

\date{\today}

\begin{abstract}

Extending the Kruppa's prescription for the continuum level density,
we have recently improved the BCS method with
seniority-type pairing force in such a way
that the effects of discretized unbound states are
properly taken into account for finite depth single-particle potentials.
In this paper, it is further shown, by employing the Woods-Saxon potential,
that the calculation of spatial observables
like nuclear radius converges as increasing the basis size
in the harmonic oscillator expansion.
Namely the disastrous problem of a ``particle gas'' surrounding nucleus
in the BCS treatment can be circumvented.
In spite of its simplicity,
the new treatment gives similar results to those
by more elaborate Hartree-Fock-Bogoliubov calculations;
e.g., it even mimics the pairing anti-halo effect.
The obtained results as well as the reason of convergence
in the new method are investigated by a variant
of the Thomas-Fermi approximation within the limited phase space
which corresponds to the harmonic oscillator basis truncation.

\end{abstract}

\pacs{21.10.Re, 21.60.Jz, 23.20.Lv, 27.70.+q}

\maketitle

%
%

\section{Introduction}
\label{sec:intro}

The advent of new radioactive beam facilities makes
it increasingly interesting to study unstable nuclei near
the neutron or proton drip line.  One of their specific features
is weak binding of constituent nucleons, which leads to
spatially extended nuclear profiles
and a striking phenomenon of the neutron halo~\cite{Tani96}.
Among many issues expected in researches of nuclei far from
the stability~\cite{DobNaz98,Dob99}, the pairing correlation
plays a special role because virtual scatterings
into continuum states occur more easily.
The basic quantity concerning the spatial distribution of nucleons
is the nuclear radius, which is also a prerequisite for the analysis
of reaction cross sections~\cite{Tani96}.
However, its theoretical evaluation in the presence of
pairing correlation is not straightforward
due to the fact that the continuum states, into which weakly bound
nucleons virtually scatter, occupy infinite volume.
In fact, the calculated radius is unreasonably large or divergent
because of finite occupation probabilities of unbound states
in the simple-minded BCS treatment;
the so-called ``neutron gas'' problem, the solution of which
requires the more sophisticated
Hartree-Fock-Bogoliubov (HFB) theory~\cite{DFT84,DNW96,DNW96a}.

A similar problem exists for the Strutinsky shell correction calculation
with finite depth potential~\cite{BFN72},
where the smoothed part of binding energy does not converge
when increasing the size of the single-particle basis:
The finite occupation probabilities of continuum states are required
for extracting the smooth part,
but the level density of unbound states is infinite,
which leads to divergence of the Strutinsky smoothed energy.
An efficient method to avoid this problem was introduced by
Kruppa~\cite{Kru98}, and was used in the shell correction method~\cite{VKN00}.
The idea of the method is to calculate the so-called continuum level density
as a difference between the level densities obtained by
diagonalizing the full Hamiltonian including a finite depth potential
and the free Hamiltonian.  By employing the harmonic oscillator
expansion, it was shown for the Woods-Saxon potential
that the smoothed energy calculated by the Kruppa method
well converges as increasing the size of basis~\cite{VKN00}.

Recently, three of the authors of this paper
have proposed a new method of BCS calculation~\cite{TST10},
which is free from the divergence as increasing the model space,
by utilizing the similar idea to Kruppa's.
In Ref.~\cite{TST10} not only the treatment
of pairing correlation but all the other procedures
in the microscopic-macroscopic method for the calculation
of nuclear binding energy are reexamined and improved.
In this paper, we further show that the new method is capable of calculating
the spatial observables like the nuclear radius without the problem
of particle gas surrounding nucleus.

\section{ Kruppa prescription for expectation values of one-body operators }
\label{sec:Kruppa}

\subsection{Basic idea}
\label{sec:basic}

In the Kruppa's prescription~\cite{Kru98}, the level density is replaced as
\begin{equation}
 g(\epsilon) \quad\Rightarrow\quad
 g\rsu{K}(\epsilon)
 =g(\epsilon)-g_0(\epsilon)\equiv
 \sum_{i=1}^{M} \delta \left( \epsilon - \epsilon_i \right)
 - \sum_{j=1}^{M} \delta ( \epsilon - \efree{j} ),
\label{eq:KruppaDensity}
\end{equation}
where $\epsilon_i$ and $\efree{j}$ are the eigenvalues of the
full and the free Hamiltonians, respectively.  In the following,
we are mainly concerned with neutrons, but it should be reminded
for protons that the Coulomb potential is included in the free Hamiltonian,
i.e., ``free'' here means that the nuclear potential is left out.
The eigenvalues in Eq.~(\ref{eq:KruppaDensity}) are calculated
by diagonalizations with the harmonic oscillator basis,
and $M$ is the total number of the basis states
commonly used in the two diagonalizations.
Both the full and free level densities,
$g(\epsilon)$ and $g_0(\epsilon)$, are divergent
as increasing the basis size ($M\rightarrow\infty$)
for the positive single-particle energy, $\epsilon > 0$.
It is shown for the Woods-Saxon potential~\cite{Kru98} that
their difference $g\rsu{K}(\epsilon)$ remains finite,
and converges, in the spherical case, to the well-known expression
in terms of the scattering phase shift $\delta_{lj}(\epsilon)$,
\begin{equation}
  \frac{1}{\pi}\sum_{lj} (2j+1)\frac{d\delta_{lj}(\epsilon)}{d\epsilon}.
\label{eq:gphaseshift}
\end{equation}
The great merit of Eq.~(\ref{eq:KruppaDensity}) is that it can
be easily applied to the deformed cases (see Ref.~\cite{Kru98}
for the corresponding expression to Eq.~(\ref{eq:gphaseshift})
in terms of the scattering S-matrix in such general cases).

Extending the idea of subtracting the free contribution
in Eq.~(\ref{eq:KruppaDensity}), we propose to calculate
the expectation value of an arbitrary one-body observable $O$
by a similar replacement as
\begin{equation}
 \langle O \rangle \quad\Rightarrow\quad
  \langle O \rangle\rsu{K}=
 \langle O \rangle - \langle O \rangle_0,
\label{eq:OKruppa}
\end{equation}
where the first (second) term is a summation of 
the expectation values with respect to
the wave functions calculated by diagonalizing the full (free)
Hamiltonian multiplied with the occupation numbers.
Note that, in the independent particle approximation,
e.g., the Hartree-Fock (HF) theory, the second term
does not contribute for bound systems, i.e., if the Fermi energy is
below the particle threshold.
The free contribution manifests itself when
the occupation probabilities of continuum states are non-zero
due to the residual interactions, e.g., in the case of
BCS treatment for pairing correlations.

Although generalizations are possible, we consider in this work
the following seniority-type separable pairing interaction,
\begin{equation}
 V\rs{pair}=-G\,\hat{P}^\dagger \hat{P},\qquad
  \hat{P}^\dagger
  =\sum_{i>0} \cutoff(\epsilon_i)\,c^\dagger_i c^\dagger_{\bar{i}},
\label{eq:SPairV}
\end{equation}
where $\bar{i}$ represents the time-reversed state of $i$, forming
a time reversal pair $(i\bar{i})$
(they are degenerate in energy, $\epsilon_i=\epsilon_{\bar{i}}$),
and $\sum_{i>0}$ means the sum is taken over all the pairs.
A cutoff of the model space is necessary for such a pairing interaction,
and it is realized by introducing a smooth cutoff
function~\cite{BFH85}, $\cutoff(\epsilon)$ (we use a different form
from that of Ref.~\cite{BFH85}), defined by
\begin{equation}
\cutoff(\epsilon) = 
 \frac{1}{2} \left[1+
 {\rm erf}\left(\frac{\epsilon-\tilde{\lambda}+\Lambda\rs{l}}{d\rs{cut}}\right)
 \right]^{1/2} \left[1+
 {\rm erf}\left(\frac{-\epsilon+\tilde{\lambda}+\Lambda\rs{u}}{d\rs{cut}}\right)
 \right]^{1/2},
\label{eq:fe}
\end{equation}
where ${\displaystyle
{\rm erf}(x)\equiv\frac{2}{\sqrt{\pi}} \int_{0}^{x}e^{-t^2}dt }$
is the error function.
We use the cutoff parameters of the pairing model space,
$\Lambda\rs{u}=\Lambda\rs{l}=1.2 \, \hbar \omega$ and
$d\rs{cut}=0.2 \,\hbar \omega$, with $\hbar\omega=41/A^{1/3}$ MeV.
The predefined parameter $\tilde{\lambda}$ in the cutoff
function~(\ref{eq:fe}) is chosen to be the smoothed Fermi energy
obtained by the Strutinsky smoothing procedure,
see Ref.~\cite{TST10} for details.

The gap equation can be derived from
the variational principle with the number constraint,
\begin{equation}
 \Delta=G\langle \hat{P}^\dagger \rangle,\qquad\mbox{and}\qquad
 N=\langle \hat{N} \rangle,
\label{eq:BCSeqFromMean}
\end{equation}
where $N$ is the neutron or proton number to be fixed.
Applying the prescription~(\ref{eq:OKruppa}), the modified gap and
number constraint equations are obtained:
\begin{eqnarray}
  \frac{2}{G} &= & \sum_{i>0}
\left[
  \frac{\cutoff (\epsilon_i)^2}{E(\epsilon_i)} 
 -\frac{\cutoff (\efree{i})^2}{E(\efree{i})} 
\right],
\label{eq:KruppaBCSGapEq}
\\
 N &=& \sum_{i>0}
\left[ 2v^2(\epsilon_i) - 2v^2(\efree{i}) \right].
\label{eq:KruppaBCSNumberEq}
\end{eqnarray}
Here the BCS quasiparticle energy and occupation probability are given,
as usual, by
\begin{equation}
 E(\epsilon)={\sqrt{(\epsilon - \lambda)^2+\cutoff(\epsilon)^2 \Delta^2}},
  \qquad
 v^2(\epsilon) = \frac{1}{2}
  \left( 1 -\frac{\epsilon-\lambda}{E(\epsilon)}\right).
\label{eq:occBCS}
\end{equation}
From these equations,
the pairing gap and the chemical potential, ($\Delta,\lambda$),
are determined for a given pairing strength $G$.
The second terms in the square brackets in Eqs.~(\ref{eq:KruppaBCSGapEq})
and~(\ref{eq:KruppaBCSNumberEq}) are the extra (negative) contributions
form the free spectra.
We call this new way of the BCS treatment Kruppa-BCS method,
and the resultant equation Kruppa-BCS equation.
The consequences of this Kruppa-BCS equation have been investigated
in detail in Ref.~\cite{TST10}.
As long as the chemical potential is well below the free particle threshold
(minimum of $\{\efree{i},i=1,2,...\}$), the Kruppa-BCS equation has
a unique solution and can be solved
in the same way as the ordinary BCS equation.
It is shown that the calculated pairing gaps
with this new method converge to reasonable values
for large basis size in contrast to those with the ordinary BCS equation,
which strongly depend on the size of basis and hardly
applicable to the calculations with large basis size~\cite{NWD94}.
See Ref.~\cite{TST10} for a comprehensive discussion.

\subsection{Root mean square radii and deformation parameters}
\label{sec:rmsdef}

\begin{figure}[ht]
\includegraphics[width=75mm]{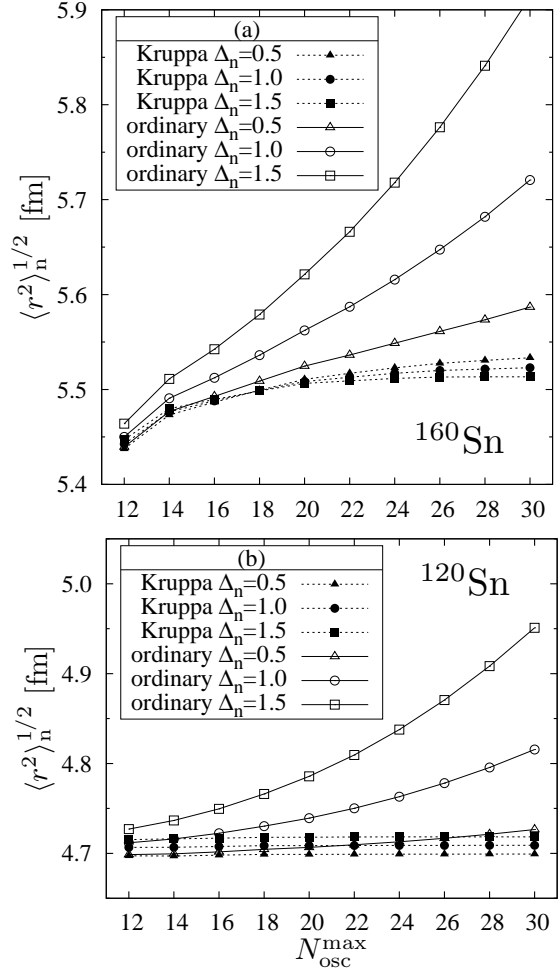}
\vspace*{-5mm}
\caption{
Neutron's rms radii as functions of the maximum harmonic
oscillator quantum number $\noscmax$, which specifies the size of basis.
Those calculated by the ordinary prescription,
$\langle r^2 \rangle\rs{n}$, and the Kruppa prescription,
$\langle r^2 \rangle\rsu{K}\rs{n}$ in Eq.~(\ref{eq:rmsK}), are included for
a near drip-line nucleus $^{160}$Sn, the panel (a),
and for a $\beta$-stable nucleus $^{120}$Sn, (b).  Both nuclei are spherical.
The neutron pairing gap is fixed to be either
$\Delta\rs{n}=0.5$, 1.0, or 1.5 MeV.
}
\label{fig:SnRMS}
\end{figure}

Once the pairing gap and the chemical potential, $(\Delta,\lambda)$,
are obtained by the Kruppa-BCS equation~(\ref{eq:KruppaBCSGapEq})
and~(\ref{eq:KruppaBCSNumberEq}), the expectation value
of one-body observables can be calculated within the BCS treatment;
for example, the root mean square (rms) radius by
\begin{equation}
 \langle r^2 \rangle\rsu{K} =
 \langle r^2 \rangle- \langle r^2 \rangle_0
 = \frac{1}{N} \sum_{i>0}\left[
 \langle i |r^2| i \rangle\, 2v^2(\epsilon_i)
 - \langle i |r^2| i \rangle_0\, 2v^2(\efree{i}) \right],
\label{eq:rmsK}
\end{equation}
where the diagonal matrix elements $\langle i |r^2| i \rangle$ and
$\langle i |r^2| i \rangle_0$ are with respect
to the Woods-Saxon and free spectra, respectively.
Figure~\ref{fig:SnRMS} shows the calculated neutron radii for a stable
nucleus $^{120}$Sn and an unstable nucleus $^{160}$Sn
in the lower and upper panels, respectively.  They are depicted
as functions of the maximum value of the harmonic oscillator quantum number,
$\noscmax$, specifying the model space.
As for the Woods-Saxon potential,
we use the parameter set recently developed by Ramon Wyss~\cite{RWyss},
which gives similar density distributions, both for neutrons and protons,
to those obtained by Skyrme and Gogny HF calculations.
In order to see the impact of the pairing correlation on the radius,
the pairing gap in this calculation is kept constant to the values
$\Delta\rs{n}=0.5$, 1.0, and 1.5 MeV, and only the number constraint
equation~(\ref{eq:KruppaBCSNumberEq}) is solved for determining
the chemical potential $\lambda\rs{n}$.
The Kruppa mean value $\langle r^2 \rangle\rsu{K}\rs{n}$
in Eq.~(\ref{eq:rmsK}) and
the ordinary mean value $\langle r^2 \rangle\rs{n}$ are compared.
Note that for the calculation of the latter, the ordinary number
equation is solved so that the chemical potentials in the calculations of
two radii, $\langle r^2 \rangle\rsu{K}\rs{n}$ and $\langle r^2 \rangle\rs{n}$,
are generally different.  As is quite evident from Fig.~\ref{fig:SnRMS},
the calculated radii by the ordinary method diverge as increasing
the basis size $\noscmax$.
The divergence is more rapid for larger pairing gaps.
There is no way to obtain
meaningful results even for the stable nucleus $^{120}$Sn.
This is nothing but the problem of the "neutron gas"
surrounding the nucleus~\cite{DFT84,DNW96,DNW96a}.
In contrast, the radii calculated by the Kruppa method converge to definite
values.  Comparing with Fig.~15 in Ref.~\cite{DFT84},
our Kruppa results nicely corresponds to those of the HFB calculation,
although the coordinate space is utilized in Ref.~\cite{DFT84}
and the box size instead of $\noscmax$
is changed to demonstrate the convergence.

It is sometimes recommended to be content
with a small model space~\cite{BFN72,MNM95},
like $\noscmax\approx 12$, for finite depth potentials,
because the plateau condition for the shell correction energy
is not met with larger model spaces.
Such a backward idea may work for stable nuclei,
but it is clear from Fig.~\ref{fig:SnRMS}~(a)
that the space $\noscmax\approx 12$ is not enough to calculate accurately
the rms radius of $^{160}$Sn, and 
using a larger model space requires inevitably the Kruppa method.

Another interesting feature of the Kruppa calculations
seen in Fig.~\ref{fig:SnRMS} is that
the radius of stable nucleus stretches for larger pairing gaps
while that of the nucleus near drip-line shrinks,
although the absolute amount of changes is very small.
The stretching of the radius due to the pairing correlation
is well-known: The pairing induces the couplings to particle states
above the Fermi surface, whose rms radii
are larger in average than those of hole states, 
and increased occupation probabilities
of particles and decreased occupation probabilities of holes
make the expectation value of radius larger
(see Fig.~\ref{fig:OBTFr}~(c) and discussions in Sec.\ref{sec:convergence}).
In spite of this effect, the radius of near drip-line nucleus
$^{160}$Sn shrinks.
We believe that this is related to the interesting
``pairing anti-halo'' effect~\cite{BDP00} in the HFB theory,
and will come back to this point later in Sec.~\ref{sec:antihalo}.

\begin{figure}[ht]
\includegraphics[width=75mm]{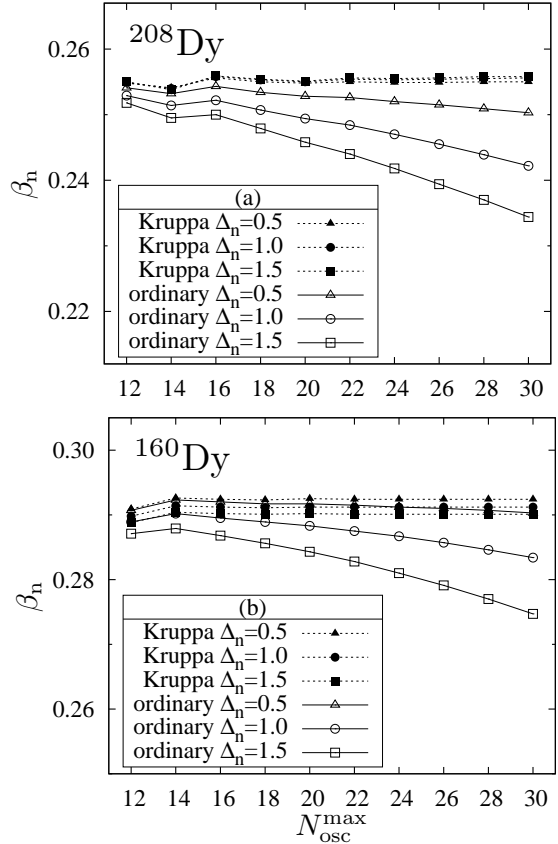}
\vspace*{-5mm}
\caption{
Neutron's deformation parameter~(\ref{eq:defbeta})
as functions of the maximum harmonic oscillator quantum number $\noscmax$.
Those calculated by the ordinary and the Kruppa prescriptions are included for
a near drip-line nucleus $^{208}$Dy, the panel (a),
and a $\beta$-stable nucleus $^{160}$Dy, (b).
Both nuclei are axially deformed.
The pairing gap, $\Delta\rs{n}=0.5$, 1.0, and 1.5 MeV, are employed.
}
\label{fig:DyBeta}
\end{figure}

In order to further
illustrate the importance of the Kruppa prescription of Eq.~(\ref{eq:OKruppa}),
presented in Fig.~\ref{fig:DyBeta} are
examples of calculation for the quadrupole deformation parameter
defined as a ratio of the quadrupole moment to the mean square radius,
\begin{equation}
 \beta = \frac{4\pi}{5}\,
 \frac{\langle r^2Y_{20} \rangle}{\langle r^2 \rangle},\qquad
 \beta\rsu{K} = \frac{4\pi}{5}\,
 \frac{\langle r^2Y_{20} \rangle\rsu{K}}{\langle r^2 \rangle\rsu{K}}
 \equiv \frac{4\pi}{5}\,
 \frac{\langle r^2Y_{20} \rangle - \langle r^2Y_{20} \rangle_0}
  {\langle r^2 \rangle - \langle r^2 \rangle_0}.
\label{eq:defbeta}
\end{equation}
In the upper and lower panels ((a) and (b)) the results for
$^{208}$Dy and $^{160}$Dy are depicted, respectively, as functions
of the basis size.  Both nuclei are axially deformed, and
the deformation parameters used in the Woods-Saxon potential are
$(\beta_2,\beta_4)=(0.244,0.053)$ and $(0.269,0.035)$, respectively,
which are obtained by the improved microscopic-macroscopic method
of Ref.~\cite{TST10}.
As in Fig.~\ref{fig:SnRMS}, the fixed pairing gaps of
$\Delta\rs{n}=0.5$, 1.0, and 1.5 MeV are used.
Apparently, the quadrupole moment as well as the radius diverges
as increasing the basis size due to the neutron gas surrounding the nucleus,
if being calculated by the ordinary BCS treatment.
One might expect that their ratio would converge to the correct value.
It is not the case, however, even in stable nuclei,
and the error is larger in neutron rich nuclei, amounting to about 10\% 
for $\Delta\rs{n}=1.5$ MeV and $\noscmax=30$ in $^{208}$Dy.
If the Kruppa prescription is used,
the results with $\noscmax \approx 12$ are already sufficiently accurate.
Namely, compared with the radius or the quadrupole moment itself,
the convergence of the deformation parameter as their ratio
against the increase of the model space is much faster.
Therefore it is better to use the Kruppa method even for
calculations with $\noscmax \approx 12$.

\subsection{Semiclassical consideration}
\label{sec:semiclassical}

In the case of the single-particle level density,
the convergence of the Kruppa density~(\ref{eq:KruppaDensity}) 
to the exact density~(\ref{eq:gphaseshift}) in the limit of
infinite model space, $M\rightarrow\infty$ (continuum limit),
can be proved rigorously~\cite{Kru98,Shl92,TO75}.
Recently, we have shown the convergence more pictorially
by employing a variant of the semiclassical (Thomas-Fermi) approximation.
We call it the oscillator-basis Thomas-Fermi (OBTF) approximation~\cite{TST10},
which is suitable to treat the problem with the truncation
in terms of the harmonic oscillator basis.
In the following, we briefly sketch this OBTF approximation, and show
that the Kruppa prescription~(\ref{eq:OKruppa})
for the expectation value of any spatial observables like the radius
is convergent as increasing the basis size, $\noscmax\rightarrow\infty$.

The basic quantity in the Thomas-Fermi approximation
is the phase space distribution function~\cite{RS80}.
Because the momentum distribution is isotropic
in the nuclear ground state~\cite{RS80},
the distribution function depends only on
the magnitude of the momentum variable, which one can transform
to the energy $\spe$.
Thus the distribution function for the single-particle Hamiltonian,
\dm{ H=\frac{\bm{p}^2}{2m}+V(\bm{r}) }, is written,
using the Heaviside step function $\theta(x)$, as
\begin{equation}
 \ftf(\bm{r},\spe)=
 \frac{(2m)^{3/2}}{2\pi^2 \hbar^3}
 \left\vert \spe - V(\bm{r}) \right\vert^{1/2}
 \theta\left( \spe - V(\bm{r}) \right),
\label{eq:lldensTF}
\end{equation}
with which the Thomas-Fermi level density can be obtained as
\dm{ \gtf(\spe)= \int \ftf(\bm{r},\spe)d^3 r }.
In the OBTF approximation, the distribution function is modified
to incorporate the effect of the limited phase space corresponding to
the truncated harmonic oscillator (HO), i.e.,
\begin{equation}
 \fobtf(\bm{r},\spe)=
 \ftf(\bm{r},\spe)\,
\theta\left(\ekmax(r)+ V(\bm{r}) -\spe \right).
\label{eq:lldensOBTF}
\end{equation}
Here the quantity $\ekmax(r)$ is the local maximum kinetic energy
of the isotropic HO potential,
\dm{ V\rs{HO}(\bm{r})=\frac{1}{2}m\omega^2 r^2 },
\begin{equation}
 \ekmax(r)= \frac{1}{2}m\omega^2\left( \rmax^2-r^2\right)
 \theta\left( \rmax-r\right),
\label{eq:ekmax}
\end{equation}
where the maximum radius $\rmax$ allowed in the HO potential is
specified by the cutoff energy $\ecut$,
related to the maximum HO quantum number $\noscmax$ of the basis truncation;
\begin{equation}
 \rmax = \sqrt{\frac{2\ecut}{m\omega^2}},\qquad
 \ecut = \hbar \omega
 \left[ (\noscmax+1)(\noscmax+2)(\noscmax+3) \right]^{1/3}.
\label{eq:rmaxecut}
\end{equation}
(Although the anisotropic HO basis is utilized in the actual calculation,
it is enough to consider the isotropic HO potential for proving
the convergence.)
Thus, the infinite model space limit $\noscmax\rightarrow\infty$
is realized by $\rmax\rightarrow\infty$ in $f_{\mbox{\tiny OB}}(\bm{r},\spe)$,
and the OBTF level density
\dm{ \gobtf(\spe)= \int \fobtf(\bm{r},\spe)d^3 r } is finite
and calculable as long as $\rmax$ (or $\noscmax$) is kept finite.
Note that this is not the case for the usual Thomas-Fermi quantities;
$\gtf(\spe)$ and $\ftf(\bm{r},\spe)$ are both infinite
for the energy $\spe$ above the particle threshold
due to the infinite spatial volume of the phase space.
In the same way, the OBTF distribution function for the free Hamiltonian,
$f^0_{\mbox{\tiny OB}}(\bm{r},\spe)$, is defined by dropping
the potential $V(\bm{r})$ (or replacing it with the Coulomb potential
for protons).  The Kruppa level density in the OBTF approximation,
$g\rsu{K}_{\mbox{\tiny OB}}(\spe)
 =g_{\mbox{\tiny OB}}(\spe)- g^0_{\mbox{\tiny OB}}(\spe)$,
is obtained by the spatial integral of
$f\rsu{K}_{\mbox{\tiny OB}}(\bm{r},\spe)
 =f_{\mbox{\tiny OB}}(\bm{r},\spe)-f^0_{\mbox{\tiny OB}}(\bm{r},\spe)$.
Both level densities, $\gobtf(\spe)$ and $\gobtf^0(\spe)$,
diverges as $\rmax\rightarrow\infty$,
but the Kruppa density $\gobtf\rsu{K}(\spe)$ remains finite for
the potential $V(\bm{r})$
that vanishes rapidly enough as $|\bm{r}|\rightarrow\infty$
like the Woods-Saxon potential.
It has been shown~\cite{TST10} that, for single-particle energies satisfying
$0 < \spe < \ekmax(\bm{r})+V(\bm{r})$ (assuming $V(\bm{r})\le 0$) everywhere,
\begin{equation}
 \gobtf\rsu{K}(\spe) \approx
 \frac{(2m)^{3/2}}{2\pi^2 \hbar^3}
 \int
 \left[\left(\spe-V(\bm{r})\right)^{1/2}-\spe^{1/2}\right]\,d^3 r,
\label{eq:gobasympt}
\end{equation}
asymptotically in the limit of $\rmax \rightarrow \infty$.

It is straightforward to apply the OBTF approximation to
the calculation of physical observables
with pairing correlation
(see, e.g.,  Ref.~\cite{RS80} for the semiclassical treatment
of the BCS theory).
For a spatial one-body observable, $O(\bm{r})$,
its Kruppa OBTF expectation value can be expressed as
\begin{equation}
 \mean{O}\rsu{K}_{\mbox{\tiny OB}}=
 \int_{-\infty}^\infty
 \gobtf\rsu{K}(\mean{O};\spe)\, 2v^2(\spe)\,d\spe,
\label{eq:OBTFKruppa}
\end{equation}
where the BCS occupation probability $v^2(\spe)$
is given in Eq.~(\ref{eq:occBCS}), and
the distribution of $\mean{O}$ with respect to the energy is defined by
\begin{equation}
 \gobtf\rsu{K}(\mean{O};\spe)\equiv
 \int O(\bm{r})\fobtf\rsu{K}(\bm{r},\spe)\,d^3 r,
\label{eq:OBTFOden}
\end{equation}
which reduces to the level density $\gobtf\rsu{K}(\spe)$
for the case $O=1$.
The problem of ``particle gas'' occurs at energies
above the particle threshold, $\spe>0$,
as in the same way as the level density.
Then, similarly to Eq.~(\ref{eq:gobasympt}),
the quantity~(\ref{eq:OBTFOden})
converges asymptotically as $\rmax \rightarrow \infty$ to
\begin{equation}
 \gobtf\rsu{K}(\mean{O};\spe) \approx
 \frac{(2m)^{3/2}}{2\pi^2 \hbar^3}
 \int
 \left[\left(\spe-V(\bm{r})\right)^{1/2}-\spe^{1/2}\right]O(\bm{r})\,d^3 r,
\label{eq:Oasympt}
\end{equation}
for the spatial observable $O(\bm{r})$
like the radius or the quadrupole moment
(assuming the upper cutoff energy
of the pairing model space by the function~(\ref{eq:fe}),
$\approx \tilde{\lambda}+\Lambda_u$, is smaller than
$\ekmax(\bm{r})+V(\bm{r})$ everywhere).
This guarantees the convergence of the Kruppa expectation value
$\langle O \rangle\rsu{K}_{\mbox{\tiny OB}}$
in the BCS treatment.

\subsection{Convergence in the Kruppa prescription}
\label{sec:convergence}

\begin{figure*}[ht]
\includegraphics[width=150mm]{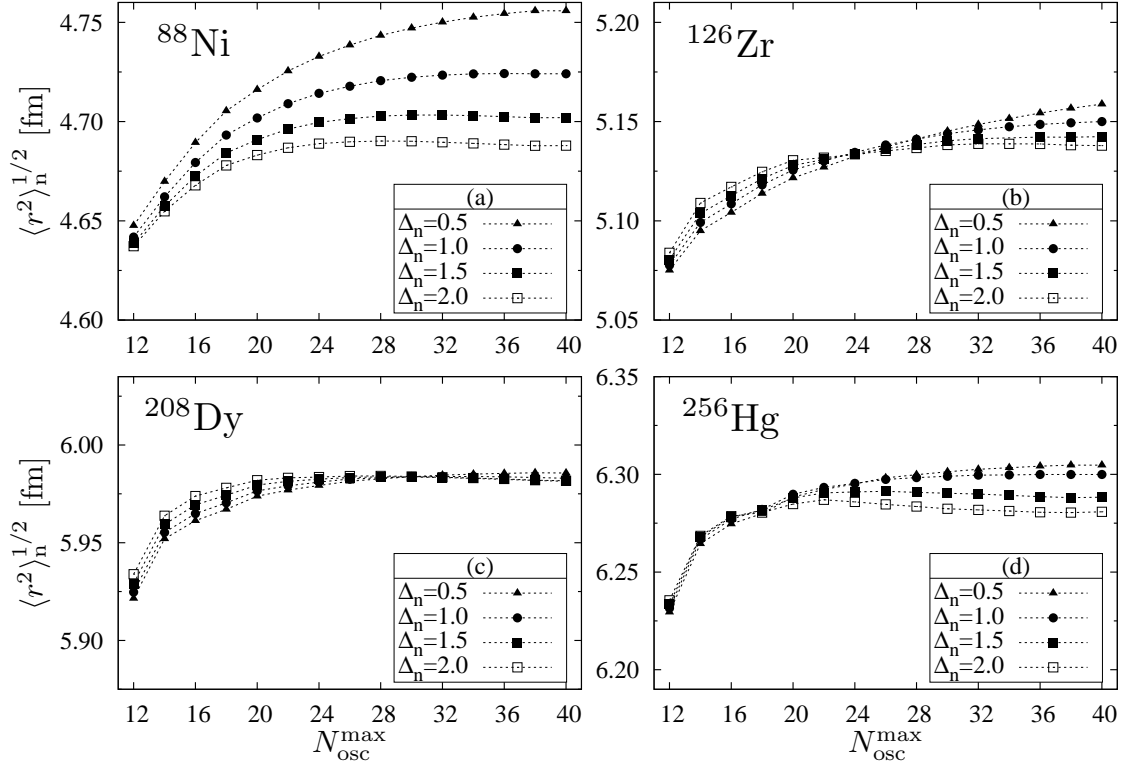}
\vspace*{-5mm}
\caption{
Convergence of neutron's rms radii calculated by the Kruppa method
as functions of $\noscmax$ for various nuclei near the drip-line,
$^{88}$Ni (a), $^{126}$Zr (b), ${208}$Dy (c), and $^{256}$Hg (d).
Only $^{208}$Dy is axially deformed.
The values of neutron pairing gap used are
$\Delta\rs{n}=0.5$, 1.0, 1.5, and 2.0 MeV.
}
\label{fig:Rconv}
\end{figure*}

In order to see how the calculated rms radii converge
as functions of $\noscmax$, we have done many test calculations~\cite{Ono10}.
Generally, the rate of convergence is fast for stable nuclei
as is already shown in Fig.~\ref{fig:SnRMS}~(b), 
while it is slow for neutron rich nuclei depending on what kind of orbits exist
near the Fermi surface.
Figure~\ref{fig:Rconv} includes four selected examples
of nuclei near the neutron drip-line
(note the extension of the abscissa to $\noscmax=40$ 
and the enlarged scale of ordinate
compared with Fig.~\ref{fig:SnRMS}).
All nuclei except $^{208}$Dy are spherical.
The radius has converged within accuracy of about 0.5\%
already at $\noscmax\approx 20$
in most cases where the pairing correlation is effective ($\Delta \ge 1$ MeV).
The very slow convergence in the weak pairing correlations, e.g., the case of
$\Delta\rs{n}=0.5$ MeV in $^{88}$Ni (panel (a))
is due to the filling of the ``halo'' orbit near the Fermi surface,
e.g., $\nu 3s_{1/2}$ (and $\nu 2d_{3/2}$) in $^{88}$Ni.
The description of such spatially extended wave functions
requires many oscillator basis states.
Another striking feature shown in Fig.~\ref{fig:Rconv}
is that the calculated radii with the largest basis size ($\noscmax=40$)
are always smaller for larger pairing gaps:
It is found that this calculated feature is rather general
for nuclei near the drip-line,
and mimics the pairing anti-halo effect in the HFB theory.
We discuss this point in more detail in Sec.~\ref{sec:antihalo}.

\begin{figure}[ht]
\includegraphics[width=75mm]{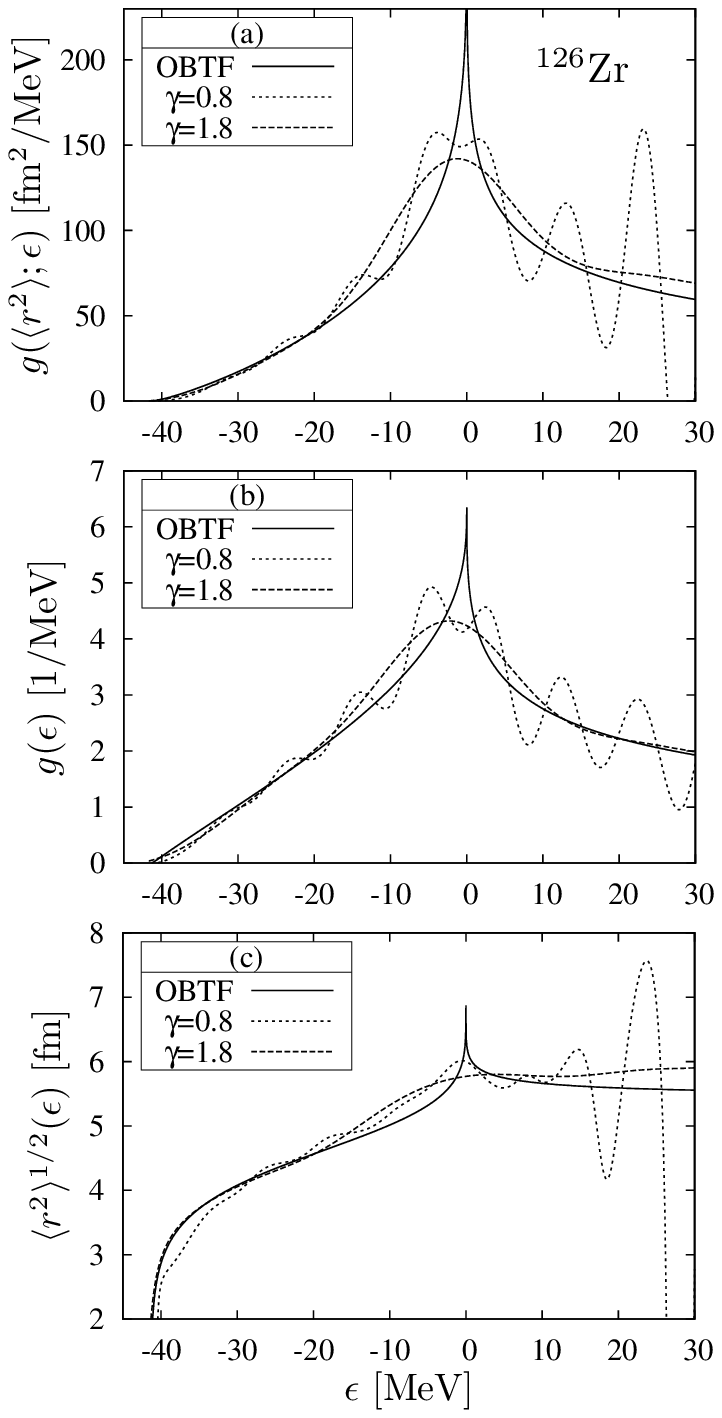}
\vspace*{-5mm}
\caption{
The distribution of $\mean{r^2}$ with respect to the single-particle
energy $\epsilon$ (c.f. Eqs.~(\ref{eq:OBTFOden}),~(\ref{eq:r2den})),
panel (a), the level density, panel (b),
and the rms radius of orbits at $\spe$
defined in Eq.~(\ref{eq:rmsorb}), panel (c),
as functions of $\epsilon$ for the spherical drip-line nucleus $^{126}$Zr.
The OBTF and Strutinsky smoothed ones with the smoothing parameter
$\gamma=0.8\,\hbar\omega$ and $1.8\,\hbar\omega$ and
with the six order of the curvature correction polynomial
($p=3$ in Ref.~\cite{TST10}) are included.
$\noscmax=40$ is used for the basis size.
}
\label{fig:OBTFr}
\end{figure}

In Ref.~\cite{TST10} the OBTF level density is compared with
the Strutinsky smoothed level density
to clarify the convergence mechanism by the Kruppa prescription,
i.e., by subtracting the contributions of free spectra.
It is instructive to perform
a similar analysis for the mean square radius.
The microscopic quantity corresponding to the distribution
of $\mean{O}=\mean{r^2}$ in Eq.~(\ref{eq:OBTFOden}) is
\begin{equation}
 g\rsu{K}(\mean{r^2};\spe)=\sum_{i>0} \left[
 \langle i |r^2| i \rangle\,\delta(\spe -\spe_i)
 - \langle i |r^2| i \rangle_0\,\delta(\spe -\efree{i})  \right],
\label{eq:r2den}
\end{equation}
with which the mean square radius in Eq.~(\ref{eq:rmsK}) is calculated by
integral
\begin{equation}
 \mean{r^2}\rsu{K}=\frac{1}{N}\int_{-\infty}^{\infty}
 g\rsu{K}(\mean{r^2};\spe)\,2v^2(\spe)\,d\spe.
\label{eq:rmsKden}
\end{equation}
In Fig.~\ref{fig:OBTFr},
we depict the Strutinsky smoothed quantity of Eq.~(\ref{eq:r2den})
and its OBTF approximation (panel (a)),
as well as the corresponding level densities (panel (b)).
The large model space with $\noscmax=40$ is used.
As for the OBTF approximation,
we use Eq.~(\ref{eq:Oasympt}) with $O=r^2$ for $\spe>0$
and the conventional Thomas-Fermi expression, i.e.,
$\left(\spe-V(r)\right)^{1/2}-\spe^{1/2}$
in the square brackets in Eq.~(\ref{eq:Oasympt}) being replaced with
$\left|\spe-V(r)\right|^{1/2}\theta(\spe-V(r))$, for $\spe<0$,
considering only the central part of the Woods-Saxon potential
with spherical symmetry for $V(r)$ \cite{TST10}.
The sharp cusp behaviors at $\spe=0$ seen
in all panels of Fig.~\ref{fig:OBTFr} are due to
the threshold effect characteristic in the OBTF approximation~\cite{TST10}.

As it was already discussed in detail in Ref.~\cite{TST10},
apart from the oscillation and the cusp at $\spe=0$,
the OBTF approximation reproduces nicely the microscopic level density
smoothed with $\gamma=1.8\,\hbar\omega$.
It also reproduces, but less nicely, the quantity $g(\mean{r^2};\spe)$.
The result of the smoothing with $\gamma=1.8\,\hbar\omega$ is larger than
the OBTF approximation in $-20$ MeV $\le \spe \le -2$ MeV 
(where the error becomes largest)
and $\spe \ge 2$ MeV
(from where there is little contribution
because of the vanishing occupation probability $v^2(\spe)$).
It is helpful to introduce a physically more meaningful quantity,
``rms radius of orbits at the single-particle energy $\spe$'',
defined by
\begin{equation}
 \mean{r^2}^{1/2}(\spe)\equiv \sqrt{g\rsu{K}(\mean{r^2};\spe)/g\rsu{K}(\spe)},
\label{eq:rmsorb}
\end{equation}
which is also depicted in Fig.~\ref{fig:OBTFr}~(c).
The corresponding quantity in the HFB theory is the
rms radius of the canonical basis states plotted versus
the expectation value of the mean field Hamiltonian,
whose gradually increasing tendency
above the particle threshold is shown in Ref.~\cite{Taj04}.
From Eqs.~(\ref{eq:gobasympt}) and~(\ref{eq:Oasympt})
(and corresponding ones in the conventional Thomas-Fermi
approximation for $\spe<0$), the Kruppa OBTF radius
$\mean{r^2}^{1/2}(\spe)$ in Eq.~(\ref{eq:rmsorb})
monotonically increases in $\spe < 0$,
takes a maximum value $\mean{r^2}^{1/2}_{\sqrt{V}}$ at $\spe=0$
and goes to an asymptotic value $\mean{r^2}^{1/2}_V$
as $\spe\rightarrow\infty$,
where (assuming the spherical symmetry)
\begin{eqnarray}
 \mean{r^2}_{\sqrt{V}}&\equiv&
 \int \sqrt{-V(r)}\,r^4dr/ \int \sqrt{-V(r)}\,r^2dr,
\label{eq:rmspotsq} \\
 \mean{r^2}_V&\equiv& \int (-V(r))\,r^4dr/ \int (-V(r))\,r^2dr.
\label{eq:rmspot}
\end{eqnarray}
The microscopic Strutinsky smoothed radii in Fig.~\ref{fig:OBTFr}~(c)
behave similarly to the OBTF expression
except a complete smearing out of the cusp (with $\gamma=1.8\,\hbar\omega$)
and an increasing overestimation at positive energies.
The monotonically increasing trend of the quantity $\mean{r^2}^{1/2}(\spe)$
in $\spe < 0$ (smoothing width $\gamma=0.8\,\hbar\omega$)
clearly indicates that the orbit with larger energy
has larger radius in average, which leads to the increase of
the total nuclear radius for larger pairing gap.

The analyses in this subsection have shown
how finite results for rms radii are obtained with the Kruppa prescription
in the OBTF approximation
and in which ways the quantum mechanical results deviate
from the results of this approximation.

\subsection{Pairing anti-halo like effect}
\label{sec:antihalo}

It is well-known that the occupation of the weakly bound orbit
with low orbital angular momentum, typically $s$-orbits, leads to
the nuclear halo and many examples are experimentally observed
in light nuclei~\cite{Tani96}.
However, it is speculated that the increased pairing correlation
binds these halo orbits more tightly and prevents
the appearance of halo phenomena in heavier nuclei~\cite{BDP00}.
To confirm this speculation is one of the most interesting subjects
both theoretically and experimentally.

\begin{figure}[ht]
\includegraphics[width=75mm]{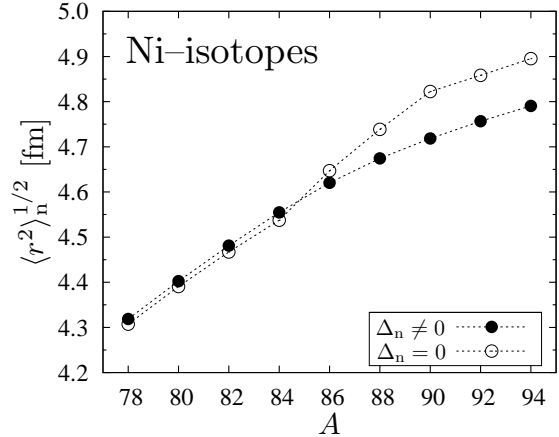}
\vspace*{-5mm}
\caption{
Neutron's rms radii calculated by the Kruppa method
with $\noscmax=40$ for the Ni--isotopes near the drip-line.
The Kruppa-BCS equation is solved to obtain the selfconsistent
pairing gap, where the pairing force strength is determined by
the improved average gap method of Ref.~\cite{TST10}.
The results without including the pairing correlation 
are also included.
}
\label{fig:NiRMS}
\end{figure}

In this respect, the Ni--isotopes are very interesting,
whose drip-line may possibly be reached in near future.
The coordinate-space HFB calculation in Ref.~\cite{GSG01},
as an example, showed that indeed an abrupt increase of radii
in the HF calculation is diminished by including the pairing correlation.

We have calculated the same quantity with the Kruppa method
in the following way.
The calculated neutron drip-line based on
the original Woods-Saxon potential~\cite{RWyss} is slightly
inside of the expected position.
Therefore, the modification of its depth
has been done according to a new systematic method developed
in Ref.~\cite{TST10}; then the drip-line isotope
is the $A=92$ nucleus (the $A=94$ nucleus is unbound),
while it is the $A=90$ nucleus in the calculation of Ref.~\cite{GSG01}.
The Kruppa-BCS equation is solved for each nucleus,
and the full microscopic-macroscopic (Strutinsky shell correction)
method is applied with the results that all the isotopes are spherical.
The average pairing gap method with $\Delta=13/\sqrt{A}$ MeV is used
to fix the strength $G$, and
the calculated pairing gaps are $\Delta\rs{n} \approx$ 1.2 to 1.7 MeV,
monotonically increasing from the $A=80$ to $A=92$ isotopes.
We depict the result of our Kruppa calculations
for radii of Ni--isotopes in Fig.~\ref{fig:NiRMS}.
Although the absolute values of radii are slightly larger in our
calculations, the behavior of how the radii increase is very
similar to the calculation shown in Fig.~7 of Ref.~\cite{GSG01}.

It may be worthwhile mentioning that a BCS treatment with including only
the resonance orbits (the resonance-BCS) is tested in Ref.~\cite{GSG01}:
The calculated radii are not good approximation
to those of HFB (no pairing anti-halo effect is reproduced),
although the obtained binding energy is very good.
This clearly indicates the subtlety of the neutron gas problem
in the conventional BCS treatment, and
we should be very careful to calculate such quantities
for weakly bound systems~\cite{DobNaz98}.

The pairing anti-halo effect in the HFB theory~\cite{BDP00}
is a result of real shrinkage of the hole component of the
quasiparticle orbits
due to the selfconsistent modification of the potentials.
Apparently, such effect is absent in the Kruppa prescription because
the BCS treatment of quasiparticle does not change the spatial
distribution of each single-particle orbit.
Then it is surprising that a similar shrinkage of the rms radius comes out
from the Kruppa-BCS calculation.  The key to understand the reason
is the rms radius $\mean{r^2}^{1/2}(\spe)$ defined in Eq.~(\ref{eq:rmsorb})
and depicted in Fig.~\ref{fig:OBTFr}~(c).
Note that the total mean square radius is calculated as an
integral~(\ref{eq:rmsKden}) with
the weight of the BCS occupation probability $v^2(\spe)$,
which is a step function at $\spe=\lambda$ smeared with
the pairing gap $\Delta$.
Therefore, if the quantity $\mean{r^2}^{1/2}(\spe)$
is larger for $\spe < \lambda$ and smaller for $\spe > \lambda$,
the increase of $\Delta$ leads to
the effective shrinkage of the total mean square radius.
In fact, this behavior, i.e., the local decrease
near the particle threshold $\spe \approx 0$,
is observed in the result of smaller smoothing width $\gamma=0.8\,\hbar\omega$
in Fig.~\ref{fig:OBTFr}~(c).

\begin{figure}[ht]
\includegraphics[width=75mm]{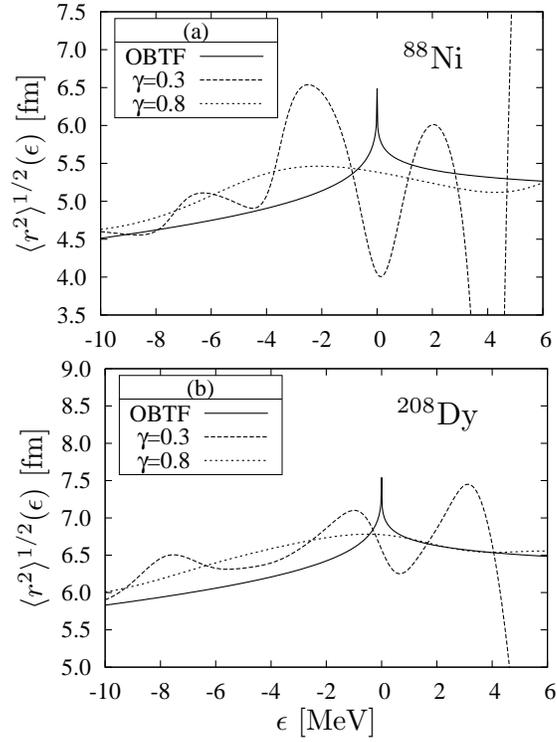}
\vspace*{-5mm}
\caption{
The same as Fig.~\ref{fig:OBTFr}~(c) but in an enlarged scale,
and for other unstable nuclei, $^{88}$Ni (a) and $^{208}$Dy (b).
The Strutinsky smoothed results with the smoothing parameter
$\gamma=0.3\,\hbar\omega$ and $0.8\,\hbar\omega$ are included.
}
\label{fig:OBTFrex}
\end{figure}

We show in Fig.~\ref{fig:OBTFrex} the same quantities
for $^{88}$Ni and $^{208}$Dy,
but are enlarged to see them more closely in the region
near the particle threshold, which is most influential for pairing
correlation in drip-line nuclei.  To show further details,
the Strutinsky smoothed results with
the smaller averaging parameter $\gamma=0.3\,\hbar\omega \approx 2\Delta$
are also included.
As is depicted in Fig.~\ref{fig:OBTFr}~(c) and Fig.~\ref{fig:OBTFrex},
we have found it quite general that the rms radius $\mean{r^2}^{1/2}(\spe)$
in Eq.~(\ref{eq:rmsorb}) is a decreasing function
near the particle threshold $\spe \approx 0$.
This trend comes from two factors: One is that, just below the threshold,
the rms radius is larger than the classical OBTF values
due to the quantum mechanical effect of the weak binding.
Another is that the rms radii of the free spectra are larger
than those of full spectra just above the threshold,
and the subtraction leads a considerably smaller total radius
than the OBTF values given in Eq.~(\ref{eq:rmspot}).
The former factor is natural but we do not understand the physical
reason of the latter factor.  Note that even if the (bound) halo orbit
exists above the Fermi level, $\lambda < \spe(\mbox{halo}) < 0$,
the subtraction effect of the free spectra is larger,
and consequently the total radius shrinks near the drip line,
$-\lambda \ltsim \Delta$.
Thus, the subtraction of the free contributions is essential
for the reduction of radius in the Kruppa method.

The actual amount of decrease strongly depends
on the nature of orbits just below the threshold;
if the orbits are of halo nature, 
the amount is larger so that the pairing shrinks the rms radius more strongly.
As is shown in Fig.~\ref{fig:Rconv},
the rms radius of $^{88}$Ni having a halo orbit $\nu 3s_{1/2}$
more strongly shrinks
than that of $^{208}$Dy when the pairing is increased,
which clearly corresponds to
the stronger decrease of the quantity $\mean{r^2}^{1/2}(\spe)$
shown in Fig.~\ref{fig:OBTFrex};
the deformation in $^{208}$Dy also prevents the appearance of
strong halo orbits due to the mixing of orbital angular momenta.
It is now clear that the mechanism of the shrinking rms radius
when increasing the pairing correlation is quite different
in our Kruppa-BCS method from that in the HFB theory.
From the discussion above, however, the essential ingredient
for the conspicuous shrinking effect is the presence of
the halo-like extended orbits below the Fermi energy,
which is common to the situation
where the strong pairing anti-halo effect is expected in the HFB approach.
In this way,
the Kruppa-BCS calculation well mimics the pairing anti-halo effect.

\subsection{Neutron skin}
\label{sec:skin}

\begin{figure}[ht]
\includegraphics[width=75mm]{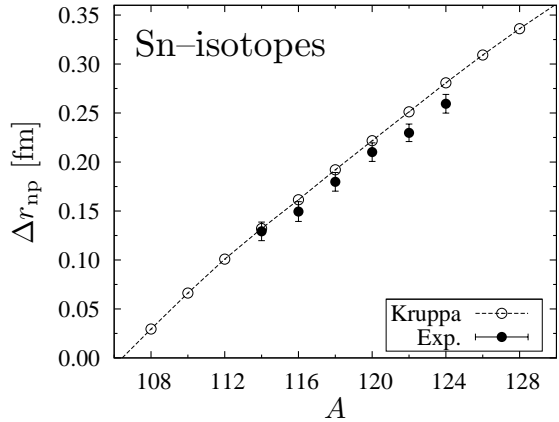}
\vspace*{-5mm}
\caption{
Neutron skin thickness calculated by the Kruppa method with $\noscmax=20$
for the Sn--isotopes.  Experimental data~\cite{KAB04} are also included.
The same calculational procedure is used as that in Fig.~\ref{fig:NiRMS}.
}
\label{fig:SnSkin}
\end{figure}

Since the radius can be accurately calculated by employing
the Kruppa prescription,
it seems also meaningful to apply it to the calculation of the neutron skin.
We compare the calculated neutron skin thickness of the Sn--isotopes
with recent experimental data~\cite{KAB04} in Fig.~\ref{fig:SnSkin}.
Here the neutron skin thickness $\Delta r\rs{np}$
is a difference of the neutron and proton rms radii,
\begin{equation}
 \Delta r\rs{np}
 =\langle r^2 \rangle\rs{n}^{1/2}-\langle r^2 \rangle\rs{p}^{1/2},
\label{eq:nskin}
\end{equation}
and they are calculated by the Kruppa method with $\noscmax=20$,
which is large enough for the (not very neutron rich) isotopes
shown in the figure.

The procedure of the calculation is the same as 
in Fig.~\ref{fig:NiRMS} for
the Ni--isotopes, see Ref.~\cite{TST10} for details.
It is interesting that the experimental data are nicely reproduced
although the Woods-Saxon potential used in this work~\cite{RWyss}
is not particularly aimed to fit the quantity like the neutron skin thickness.
(However, there are considerable ambiguities
in the experimental neutron skin thickness depending on
which types of experiments are used to extract it~\cite{KFA99,TJL01}.)
For the Sn--isotopes, a recent Skyrme HF+BCS calculation
with the SLy4 functional gives slightly smaller neutron skin thickness,
see, e.g., Ref.~\cite{SGG07}.
It is pointed out that the neutron skin has strong correlations
with the coefficient of the asymmetry energy in the mass formula
and with the equation of state of the asymmetric nuclear matter,
see, e.g., Refs.~\cite{Fur02,WVR09}.
The determination of skin thickness is therefore very important
both theoretically and experimentally.
We hope that one can utilize the Kruppa method used in this work as an
equal alternative for such an investigation if one optimizes the
parameters of the Woods-Saxon potential as well as the liquid drop model.

\subsection{Other observables}
\label{sec:otherO}

As an example of different kinds of observables,
we consider the moment of inertia
about the $x$-axis (perpendicular to the symmetry axis)
employing the Kruppa prescription,
\begin{equation}
 \cJ\rsu{K}=\cJ-\cJ_0.
\label{eq:cJKruppa}
\end{equation}
Here the free contribution, for example, is given as~\cite{RS80}
\begin{equation}
 \cJ_0=2\sum_{i,j>0}
 \frac{\left|\langle i|J_x|j\rangle_0\right|^2}{E(\efree{i})+E(\efree{j})}
 \left(u(\efree{i})v(\efree{j})-u(\efree{j})v(\efree{i})\right)^2,
\label{eq:cJfree}
\end{equation}
with $\langle i|J_x|j\rangle_0$ being the matrix element of
the operator $J_x$ with respect to the free spectra,
and $u(\epsilon)=\sqrt{1-v(\epsilon)^2}$.
Eqs.~(\ref{eq:cJKruppa}) and (\ref{eq:cJfree}) are
obtained by the cranking method
with taking the limit of zero rotational frequency:
\begin{equation}
 \cJ\rsu{K}=
 \mathop{\mbox{lim}}_{\omega\rs{rot}\rightarrow 0}
 \frac{\langle J_x \rangle\rsu{K}}{\omega\rs{rot}}
 =\mathop{\mbox{lim}}_{\omega\rs{rot}\rightarrow 0}
 \left[
 \frac{\langle\omega\rs{rot}|J_x|\omega\rs{rot}\rangle}{\omega\rs{rot}}
 -\frac{\langle\omega\rs{rot}|J_x|\omega\rs{rot}\rangle_0}{\omega\rs{rot}}
 \right],
\label{eq:crankKruppa}
\end{equation}
where the state $|\omega\rs{rot}\rangle$ is the cranked mean field
associated with $H'=H-\omega\rs{rot}J_x$,
and $|\omega\rs{rot}\rangle_0$ is the same but for the free Hamiltonian.
Thus, the moment of inertia is essentially the expectation value
of the angular momentum operator,
which is composed not only of the coordinate
but also of the momentum variables.

\begin{figure}[ht]
\includegraphics[width=75mm]{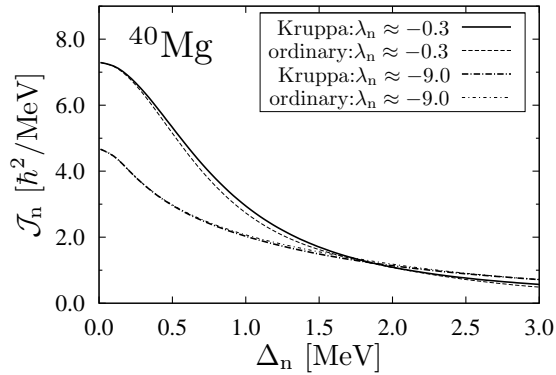}
\vspace*{-5mm}
\caption{
Neutron's contributions to moment of inertia
as functions of the neutron pairing gap
calculated by the Kruppa and ordinary methods
for the drip-line nucleus $^{40}$Mg.
The size of model space
is specified by $\noscmax=30$, with which the results well converge.
In the same way as Ref.~\cite{YS08}, are included two types of calculations,
one for a weakly bound situation, $\lambda\rs{n}\approx -0.3$ MeV,
and another for a deeply bound situation,
$\lambda\rs{n}\approx -9.0$ MeV.
The original depth of the Woods-Saxon potential is modified
in each case. See text for details.
}
\label{fig:JmMg}
\end{figure}

The effect of weak binding of constituent nucleons on the moment of inertia
in neutron rich nuclei has been investigated in Ref.~\cite{YS08}
by the coordinate-space HFB method (the particle-hole channel is
approximated by using the Woods-Saxon potential).
It has been reported that the dependence of moment of inertia
on the pairing gap is much stronger in drip-line nuclei, e.g., in $^{40}$Mg.
We have done similar calculations but by using the simple BCS
with the Kruppa prescription~(\ref{eq:cJKruppa}) and without it.
The results of neutron moment of inertia, $\cJ\rs{n}$,
are depicted in Fig.~\ref{fig:JmMg}.
The used deformation parameters are $\beta_2=0.3$ ($\beta_4=0.0$)~\cite{YS08}.
Two calculations are presented in Ref.~\cite{YS08}
with the neutron chemical potential being $\lambda\rs{n}\approx -0.3$
and $-9.0$ MeV at $\Delta\rs{n}\approx 0$,
the latter of which corresponds to a stable nucleus and
is artificially realized by deepening the potential.
We have done the same adjustment to match with these calculations;
the depth of the original Woods-Saxon potential~\cite{RWyss} is modified
by multiplying a factor 0.986 (1.365) for obtaining the desired value
$\lambda\rs{n}\approx -0.3$ ($-9.0$) MeV.
Our result is very similar to that shown in Fig.~11 of Ref.~\cite{YS08};
it reproduces
the feature of stronger dependence of moment of inertia on pairing correlation
for weakly bound systems.
However, the free contribution $\cJ_0$ is rather small.
In fact, the result $\cJ\rs{n}\rsu{K}>\cJ\rs{n}$ for the
weak binding situation ($\lambda\rs{n}\approx -0.3$ MeV)
might seem strange at first sight
because $\cJ_0$ in Eq.~(\ref{eq:cJKruppa}) is always positive.
Note that the Kruppa-BCS equation is solved for
the calculation of $\cJ\rsu{K}$ instead of the ordinary BCS equation for $\cJ$.
Then the chemical potential obtained by the Kruppa-BCS equation is nearer
to the particle threshold, and the moment of inertia $\cJ$ becomes larger.
The difference, $\cJ$($\lambda$(Kruppa-BCS))$-\cJ$($\lambda$(ordinary-BCS)),
is larger than $\cJ_0$, leading to the final result, $\cJ\rsu{K}>\cJ$.
This clearly shows that even if ${\cJ}_0>0$ its contribution is minor.
In fact, for the stable (deeply bound) situation
($\lambda\rs{n}\approx -9.0$ MeV)
the difference between the Kruppa and ordinary calculations
is negligible as is seen in Fig.~\ref{fig:JmMg}.
We have done systematic calculations
for many heavier nuclei~\cite{Ono10} and this is generally the case.
Namely, the neutron gas problem is not so harmful as far as the calculation
of moment of inertia is concerned; subtraction of the free contribution
in Eq.~(\ref{eq:cJKruppa}) is not necessarily required.
This is mainly because the free matrix elements of the $J_x$ operator
in Eq.~(\ref{eq:cJfree}) are small, which reflects that
the momentum operators contained
in the angular momentum~($\bm{l} = \bm{r} \times \bm{p}$)
do not favor spatially-extended free wave functions
in contrast to the spatial observable like the radius~($\bm{r}^2$).

\section{Summary}
\label{sec:summary}

We have extended the original Kruppa prescription
for the single-particle level density to the calculation of one-body
spatial observables like nuclear radius in the presence of pairing correlations.
By simply subtracting the contribution of the free spectra,
the effects of continuum states can be properly taken into account.
The convergence property as increasing the basis size is carefully examined
both numerically and analytically, and it is confirmed that
reasonable convergence is attained for radius with manageable basis sizes
except for the case of the halo-like situation with weak pairing correlation.

The results of a few applications of the Kruppa method are presented
for the basic quantities like nuclear radii, quadrupole moments,
the neutron skin thickness, and the moment of inertia
in very neutron rich nuclei.
As for the moment of inertia, which is essentially the expectation
value of the angular momentum operator at finite rotational frequency
and is considered to be an example of operators consisting of both
the coordinate and momentum,
the effect of subtracting the free contributions is small.
It has been found that the pairing anti-halo like effect
naturally emerges, although the underlying mechanism
is quite different.

In this way, it has been shown that
the so-called neutron gas problem is circumvented,
and the new Kruppa-BCS method gives
reliable results similar to those of more sophisticated HFB calculations.
Combined with the improved microscopic-macroscopic method
developed in Ref.~\cite{TST10},
we hope that the Kruppa prescription provides a simple, yet useful,
new method for investigating nuclei far from stability.

For the continuum single-particle level density,
the Kruppa prescription in Eq.~(\ref{eq:KruppaDensity})
has a sound basis~\cite{Kru98,Shl92,TO75}.
However, the replacement in Eq.~(\ref{eq:OKruppa}) proposed in this work
is missing such a solid foundation.
It is an interesting future subject
to clarify what is the meaning of this replacement,
or on what kind of approximations it is based,
from more fundamental many-body theory.

\section*{ACKNOWLEDGEMENTS}

Useful discussions with Masayuki~Yamagami are greatly appreciated.
This work is supported in part by the JSPS Core-to-Core Program,
International Research Network for Exotic Femto Systems (EFES),
and by Grant-in-Aid for Scientific Research (C) 
No.~18540258 from Japan Society for the Promotion of Science.
%

\vspace*{10mm}


\end{document}